\documentclass[prd,preprint,showpacs,floatfix,superscriptaddress]{revtex4}
\usepackage{amsmath}
\usepackage{latexsym}
\usepackage{amsfonts}
\usepackage{graphicx}
\usepackage{bm}
\usepackage{subfigure}

\begin{document}

\title{Probing the cosmic acceleration from combinations of different data sets}

\author{Yungui Gong}
\email{gongyg@cqupt.edu.cn}
\affiliation{College of Mathematics and Physics,
Chongqing University of Posts and Telecommunications, Chongqing 400065, China}

\author{Bin Wang}
\email{wangb@fudan.edu.cn}
\affiliation{College of Mathematics and Physics, Chongqing University of Posts and Telecommunications, Chongqing 400065, China}
\affiliation{Department of Physics, Fudan
University, Shanghai 200433, China}
\author{Rong-Gen Cai}
\email{cairg@itp.ac.cn}
\affiliation{College of
Mathematics and Physics, Chongqing University of Posts and
Telecommunications, Chongqing 400065, China}
\affiliation{Institute of Theoretical Physics, Chinese Academy of Sciences,\\
Beijing 100190, China}
\preprint{1001.0807}


\begin{abstract}
We examine in some detail the influence of the systematics in
different data sets including type Ia supernova sample, baryon
acoustic oscillation data and the cosmic microwave background
information on the fitting results of the Chevallier-Polarski-Linder
parametrization. We find that the systematics in the data sets does
influence the fitting results and leads to different evolutional
behavior of dark energy. To check the versatility of
Chevallier-Polarski-Linder parametrization, we also perform the
analysis on the Wetterich parametrization of dark energy. The
results show that both the parametrization of dark energy and the
systematics in data sets influence the evolutional behavior of dark
energy.

\end{abstract}


\maketitle

\section{Introduction}

The convincing fact that our universe is experiencing accelerated expansion \cite{acc1,acc2}
has become one of the most important and mysterious
issues of modern cosmology. In the framework of general relativity, the cosmic acceleration
is attributed to an exotic energy called dark energy
(DE). The simplest candidate of the DE is the cosmological constant with equation of state
 (EOS) $w=-1$. Despite some severe problems such as the
cosmological constant problem and the coincidence problem, the cosmological constant is
 favored by current astronomical observations. Besides,
there are some other DE models with EOS varying with cosmic time either above $-1$ or
below $-1$. Interestingly, some current observations even
give the hint that EOS of DE has crossed $-1$ at least once \cite{gong07,gong}. To narrow
 down the DE candidates, one has to examine the EOS carefully by
confronting different observational data sets including the type Ia supernova (SNIa) luminosity
distance,
the cosmic microwave background (CMB) temperature anisotropy and the
baryon acoustic oscillations (BAO) in the galaxy power spectrum, and this could be the best
what we can do to approach the truth of DE.

In addition to the EOS, a new diagnostic of DE was introduced in
\cite{sahni} which is called  $Om$ diagnostic. It is a combination
of the Hubble parameter and the cosmological redshift, which depends
upon the first derivative of the luminosity distance and is less
sensitive to observational errors than the EOS. If the value of $Om$
is the same at different redshifts, then the DE is the cosmological
constant. The slope of $Om$ can differentiate between different DE
models with $w>-1$ or  $w<-1$  even if the value of the matter
density is not accurately known.

Analyzing the Constitution SNIa sample \cite{consta} together with
the BAO data \cite{sdss6,sdss7} by using the popular
Chevallier-Polarski-Linder (CPL) parametrization
\cite{cpl1,linder03}, it was revealed that there appears the
increase of $Om$ and $w$ at redshifts $z<0.3$ \cite{star}. This
suggests that the cosmic acceleration may have already been over the
peak and now the acceleration is slowing down. However, including
the CMB shift parameter which is another independent observable at
high redshift, it was found that the result changes dramatically and
the value of $Om$ becomes un-evolving which is consistent with the
$\Lambda$CDM model. Further check shows that the CPL ansatz is
unable to fit the data simultaneously at low and high redshifts. It
was argued that this could either  due to the systematics in some of
the data sets which is not sufficiently understood or because the
CPL parametrization is not versatile to accommodate the cosmological
evolution of DE suggested by the data \cite{star}. The data sets
used in the analysis of \cite{star} are limited to the Constitution
set of 397 SNIa in combination with BAO distance ratio of the
distance measurements obtained at $z=0.2$ and $z=0.35$ and the CMB
shift parameter. It is expected that if more combinations of new
data sets are included, these arguments can be clarified. In
\cite{gong}, it was showed that the result of the analysis on the
CPL model \cite{star} heavily depends on the choice of BAO data. The
result obtained by using the BAO distance ratio data was found not
consistent with that by using other observational data, and this
inconsistency can be overcome if the BAO $A$ parameter \cite{baoa}
is employed instead \cite{gong}. In this work we are going to
investigate this problem further by comparing different data set
combinations among SNIa, BAO and CMB.

\section{Observational data}

For the SNIa data, we use the Constitution sample \cite{consta} and the first year
Sloan digital sky survey-II (SDSS-II) SNIa (hereafter Sdss2)
\cite{sdss2}. The Constitution sample consists of the Union sample \cite{union}
together with 185 CfA3 SNIa data,
which totally contains 397 SNIa. The CfA3 addition makes the cosmologically
useful sample of nearby SNIa much larger than before, which reduces the statistical
uncertainty to the point where systematics plays the largest role. To test the systematic
 differences and consistencies, in \cite{consta} four
light curve fitters, $SALT$, $SALT2$, $MLCS2k2$ with $R_V = 3.1$ (MLCS31), $MLCS2k2$ with
$R_V = 1.7$ (MLCS17), have been used. For the
Constitution SNIa data using the template SALT (hereafter Csta), the intrinsic uncertainty
of 0.138 mag for each CfA3 SNIa, the peculiar velocity
uncertainty of $400$km/s, and the redshift uncertainty have been considered \cite{consta}.
These data were suggested by observers to be the best data for
model independent analysis of the expansion history.
The Constitution SNIa data using the template SALT2
(hereafter Cstb), excludes the SNIa with $z<0.01$ or $t_{1st}>10d$, so it has 351 SNIa data.
 Using the template  MLCS17 on the Constitution data
(hereafter Cstc), the SNIa with $A_v\geq 1.5$ and $t_{1st}>10d$  has
been cut out  and it excludes those SNIa whose MLCS17 fit has a
reduced $\chi^2_\nu$ being 1.6 or higher, thus Cstc has only 372
SNIa data. The Cstd sample is formed by using  the template  MLCS31
on the Constitution sample  and cutting out the SNIa with $A_v\geq
1.5$ and $t_{1st}>10d$. It excludes any SNIa whose MLCS31 fit has a
reduced $\chi^2_\nu$ being 1.5 or higher, so that it contains 366
SNIa data. We will examine the systematic differences brought
 by the Constitution sample  with different light
curve fitters.
In addition we will also consider Sdss2 sample and
investigate the systematic difference it brought,  to compare with the
Constitution sample. The Sdss2 consists of 103 new SNIa with
redshifts $0.04<z<0.42$, discovered during the first season of the
SDSS-II supernova survey, 33 nearby SNIa with redshfits
$0.02<z<0.1$, 56 SNIa with redshifts $0.16<z<0.69$ from ESSENCE
\cite{riess,essence}, 62 with redshifts $0.25<z<1.01$ SNIa from SNLS
\cite{snls}, and 34 SNIa with redshifts $0.21<z<1.55$ from HST. For
simplicity,
 we only use the Sdss2 data with the template
MLCS2K2. For the 288 Sdss2 SNIa data, we also consider the redshift uncertainties from spectroscopic
 measurements and from peculiar motions of
the host galaxy. For the redshift uncertainty, $\sigma_{z,pec}$, arises from peculiar velocities of
and within host galaxies, we take
$\sigma_{z,pec} = 0.0012$. In addition, the intrinsic error of 0.16 mag is added to the uncertainty
of the distance modulus \cite{sdss2}.

To fit the SNIa data, we define
\begin{equation}
\label{chi1} \chi^2(\bm{p},H_0)=\sum_{i=1}\frac{[\mu_{obs}(z_i)-\mu(z_i,\bm{p},H_0)]^2}{\sigma^2_i},
\end{equation}
where the extinction-corrected distance modulus $\mu(z,\bm{p},H_0)=5\log_{10}[d_L(z)/{\rm Mpc}]+25$,
$\sigma_i$ is the total uncertainty in the
SNIa data, and the luminosity distance for a flat universe is
\begin{equation}
\label{lum} d_L(z,\bm{p},H_0)=\frac{1+z}{H_0} \int_0^z \frac{dx}{E(x)},
\end{equation}
where the dimensionless Hubble parameter $E(z)=H(z)/H_0$. Because
the normalization of the luminosity distance is unknown, the
nuisance parameter $H_0$ in the SNIa data is not the observed Hubble
constant. We marginalize over the nuisance parameter $H_0$ with a
flat prior, which leads to~\cite{gong08}
\begin{equation}
\label{chi} \chi^2_{sn}(\bm{p})=\sum_{i=1}\frac{\alpha_i^2}{\sigma^2_i}-\frac{(\sum_i\alpha_i/\sigma_i^2
-\ln 10/5)^2}{\sum_i 1/\sigma_i^2}
-2\ln\left(\frac{\ln 10}{5}\sqrt{\frac{2\pi}{\sum_i 1/\sigma_i^2}}\right),
\end{equation}
where $\alpha_i=\mu_{obs}(z_i)-25-5\log_{10}[H_0 d_L(z_i)]$.

Besides considering these SNIa data sets individually, our analysis
will also investigate the combination of SNIa data with the BAO
data. For the BAO data, we first use the
BAO distance measurements obtained at $z=0.2$ and $z=0.35$ from joint analysis
of the 2dFGRS and SDSS data \cite{sdss7}. Defining
$d_z(\bm{p},H_0)=r_s(z_d)/D_V(z)$, where the comoving sound horizon and effective distance are
\begin{equation}
\label{rshordef} r_s(z,\bm{p})=\int_z^\infty \frac{dx}{c_s(x)E(x)},
\end{equation}
\begin{equation}
\label{dvdef} D_V(z,\bm{p},H_0)=\left[\frac{d_L^2(z)}{(1+z)^2}\frac{z}{H(z)}\right]^{1/3}
=H_0^{-1}\left[\frac{z}{E(z)}\left(\int_0^z\frac{dx}{E(x)}\right)^2\right]^{1/3}.
\end{equation}
The redshift $z_d$ at the baryon drag epoch is fitted with the formula~\cite{hu98}
\begin{equation}
\label{zdfiteq} z_d=\frac{1291(\Omega_{m0} h^2)^{0.251}}{1+0.659(\Omega_{m0} h^2)^{0.828}}[1+b_1(\Omega_b h^2)^{b_2}],
\end{equation}
\begin{equation}
\label{b1eq} b_1=0.313(\Omega_{m0} h^2)^{-0.419}[1+0.607(\Omega_{m0} h^2)^{0.674}],\quad b_2=0.238
(\Omega_{m0} h^2)^{0.223},
\end{equation}
where $\Omega_{m0}$ is the current value of the dimensionless matter energy density,
$\Omega_b$ is the dimensionless
baryon matter energy density, and $h=H_0/100$.
The sound speed $c_s(z)=1/\sqrt{3[1+\bar{R_b}/(1+z)}]$, where $\bar{R_b}=315000\Omega_b h^2(T_{cmb}/2.7{\rm K})^{-4}$
and $T_{cmb}=2.726$K. In \cite{sdss7}, two
points $d_{0.2}$ and $d_{0.35}$ and their covariance matrix are given. We will use
 $d_{0.2}=0.1905\pm 0.0061$, $d_{0.35}=0.1097\pm 0.0036$ and
their covariance matrix (hereafter Bao2) as the second sample of BAO data in the data fitting. When we use these data,
 we add two more parameters
$\Omega_b h^2$ and $h$. Based on these two BAO distance measurements, we can derive the BAO distance
ratio
$D_V(0.35)/D_V(0.2)=1.736\pm 0.065$ (hereafter BaoR), which is relatively model independent quantity. This BAO
 distance ratio was employed in the
analysis in \cite{star}. In our case we will further use another two radial BAO data at redshifts $z=0.24$ and $z=0.43$
\cite{baoz} by using $\Delta
z(z)=H(z)r_s(z_d)/c$ (hereafter BaoZ). Combining BAO data sets Bao2 and BaoZ, we have the data set called Bao4.
Therefore, for different BAO data sets, we can perform $\chi^2$ statistics for the model
parameter $\bm{p}$ as follows
\begin{equation}
\label{baoR} \chi^2_{Baor}(\bm{p})=\frac{[D_V(0.35)/D_V(0.2)-1.736]^2}{0.065^2},
\end{equation}
\begin{equation}
\label{baoz} \chi^2_{BaoZ}(\bm{p},h,\Omega_b h^2)=\frac{[\Delta z(0.24)-0.0407]^2}{0.0011^2}+\frac{[\Delta z(0.43)-0.0442]^2}{0.0015^2},
\end{equation}
and
\begin{equation}
\label{bao2} \chi^2_{Bao2}(\bm{p},h,\Omega_b h^2)=\sum_{ij}\Delta d_i{\rm Cov}^{-1}(d_i,d_j)\Delta d_j,
\end{equation}
where $i,j=0.2,0.35$, $\Delta d_{0.2}=d_{0.2}-0.1905$, $\Delta d_{0.35}=d_{0.35}-0.1097$.

Since the SNIa and BAO data contain information about the universe
at relatively low redshifts, we will  include the CMB information by
implementing the Wilkinson microwave anisotropy probe 5 year (WMAP5)
data to probe the entire expansion history up to the last scattering
surface. In addition to employing the CMB shift parameter $R$
 by defining the reduced distance in the flat universe to the last scattering surface $z_*$ as
 done in \cite{star}
\begin{equation}
\label{shift} R(\bm{p})=\sqrt{\Omega_{m0}}\int_0^{z_*}\frac{dz}{E(z)}=1.71\pm 0.019,
\end{equation}
we will add the acoustic scale $l_a$ in the data analysis. The
acoustic scale $l_a$ is defined by
\begin{equation}
\label{ladefeq} l_a=\frac{\pi d_L(z_*)}{(1+z_*)r_s(z_*)}=302.1\pm
0.86,
\end{equation}
where the redshift $z_*$ is given by \cite{hu96}
\begin{equation}
\label{zstareq} z_*=1048[1+0.00124(\Omega_b h^2)^{-0.738}][1+g_1(\Omega_{m0} h^2)^{g_2}]=1090.04\pm 0.93,
\end{equation}
\begin{equation}
g_1=\frac{0.0783(\Omega_b h^2)^{-0.238}}{1+39.5(\Omega_b h^2)^{0.763}},\quad g_2=\frac{0.560}{1+21.1
(\Omega_b h^2)^{1.81}}.
\end{equation}
We have three fitting parameters $x_i=(R,\ l_a,\ z_*)$ to include
the CMB information now. Thus we have
 $\chi^2_{cmb}(\bm{p},h,\Omega_b
h^2)=\sum_{ij}\Delta x_i {\rm Cov}^{-1}(x_i,x_j)\Delta x_j$, $\Delta x_i=x_i-x_i^{obs}$ and Cov$(x_i,x_j)$ is
 the covariance matrix for the three
parameters \cite{wmap5}. The CMB information provides a systematic
check of the DE model by combining with the low redshift SNIa and
BAO data sets.

\section{Fitting Results}

We employ the Monte-Carlo Markov Chain (MCMC) method to explore the
parameter space $\bm{p}$ and the nuisance parameters $h$ and
$\Omega_b h^2$ in the data analysis.  Our MCMC code \cite{gong08} is based on the
publicly available package COSMOMC \cite{cosmomc}.

We  first explore the widely used CPL parametrization
\cite{cpl1,linder03}
\begin{equation}
\label{cplwz}
w(z)=w_0+\frac{w_a \, z}{1+z}.
\end{equation}
In this model, the dimensionless Hubble parameter for a flat
universe is
\begin{equation}
\label{cplez} E^2(z)=\frac{H^2(z)}{H^2_0}=\Omega_{m0}(1+z)^3+(1-\Omega_{m0})(1+z)^{3(1+w_0+w_a)}\exp(-3w_a z/(1+z)).
\end{equation}
In this model, we have three parameters $\bm{p}=(\Omega_{m0},\ w_0,\ w_a)$. Knowing the expansion history of the universe, we can construct the
$Om$ diagnostic by defining
\begin{equation}
\label{omzeq} Om(z)=\frac{E^2(z)-1}{(1+z)^3-1}.
\end{equation}
$Om$ diagnostic is useful in establishing the properties of DE at low redshifts. The constant $Om$ indicates that
the DE is the cosmological
constant and the bigger value of $Om$ shows that $w$ is bigger \cite{sahni}.
$Om(z)$ is less sensitive to observational errors than EOS $w(z)$. Due
to the degeneracy between $\Omega_{m0}$ and $w$, the property of $w(z)$ depends on the current value
of matter energy density $\Omega_{m0}$. However, $Om$ diagnostic provides a null test
of the cosmological constant without knowing the exact value of $\Omega_{m0}$.
To reconstruct $Om(z)$, we need to apply the specific model and model parameters. Following
\cite{star}, we consider the uncertainties of $\Omega_{m0}$, $w_0$ and $w_a$ when we we reconstruct $Om(z)$.

\begin{figure}[t]
$\begin{array}{c}
\subfigure[]{\includegraphics[width=3in]{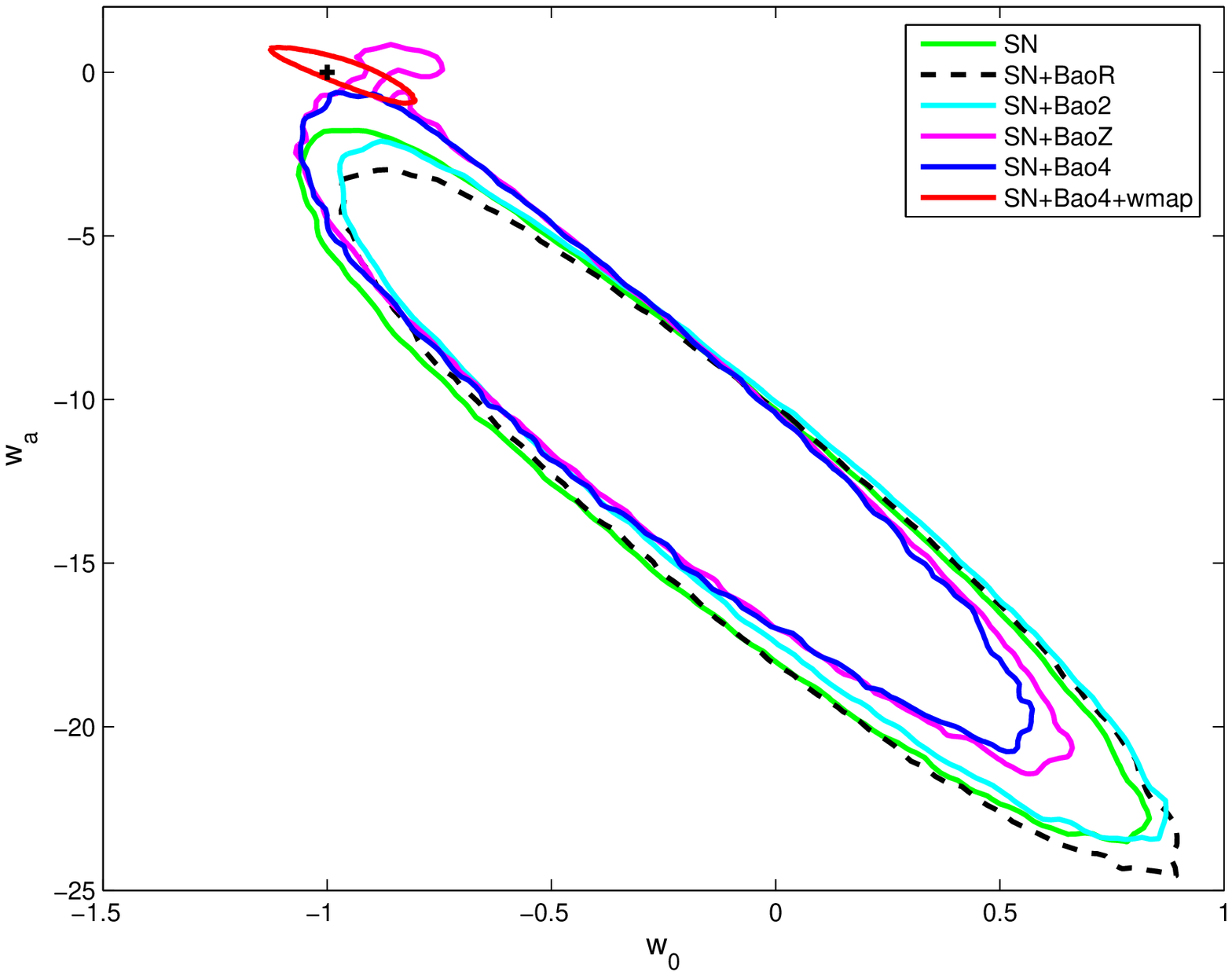}
\label{cstadcpl:a}}
\subfigure[]{ \includegraphics[width=3in]{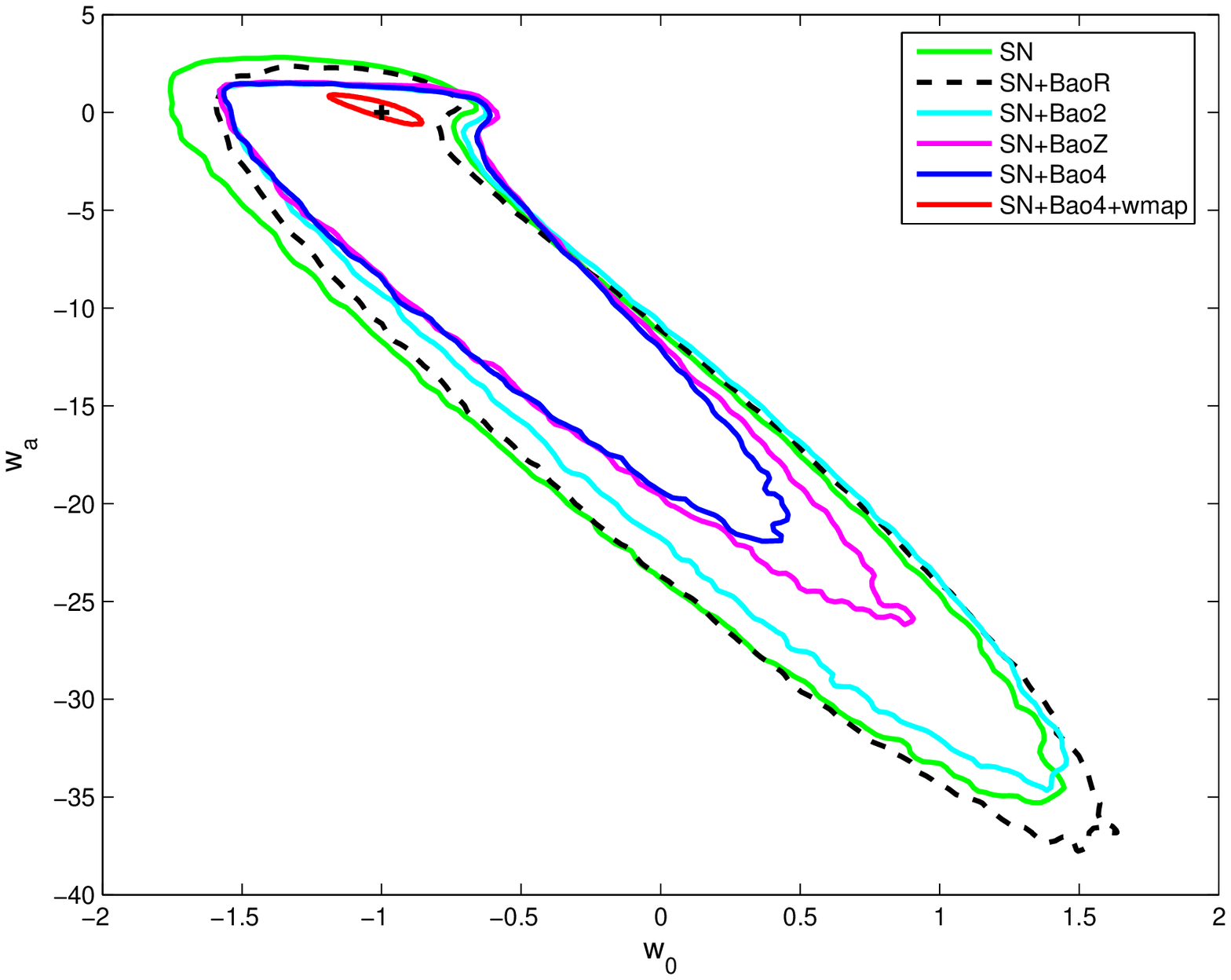}
\label{cstadcpl:b}}\\
\subfigure[]{\includegraphics[width=3in]{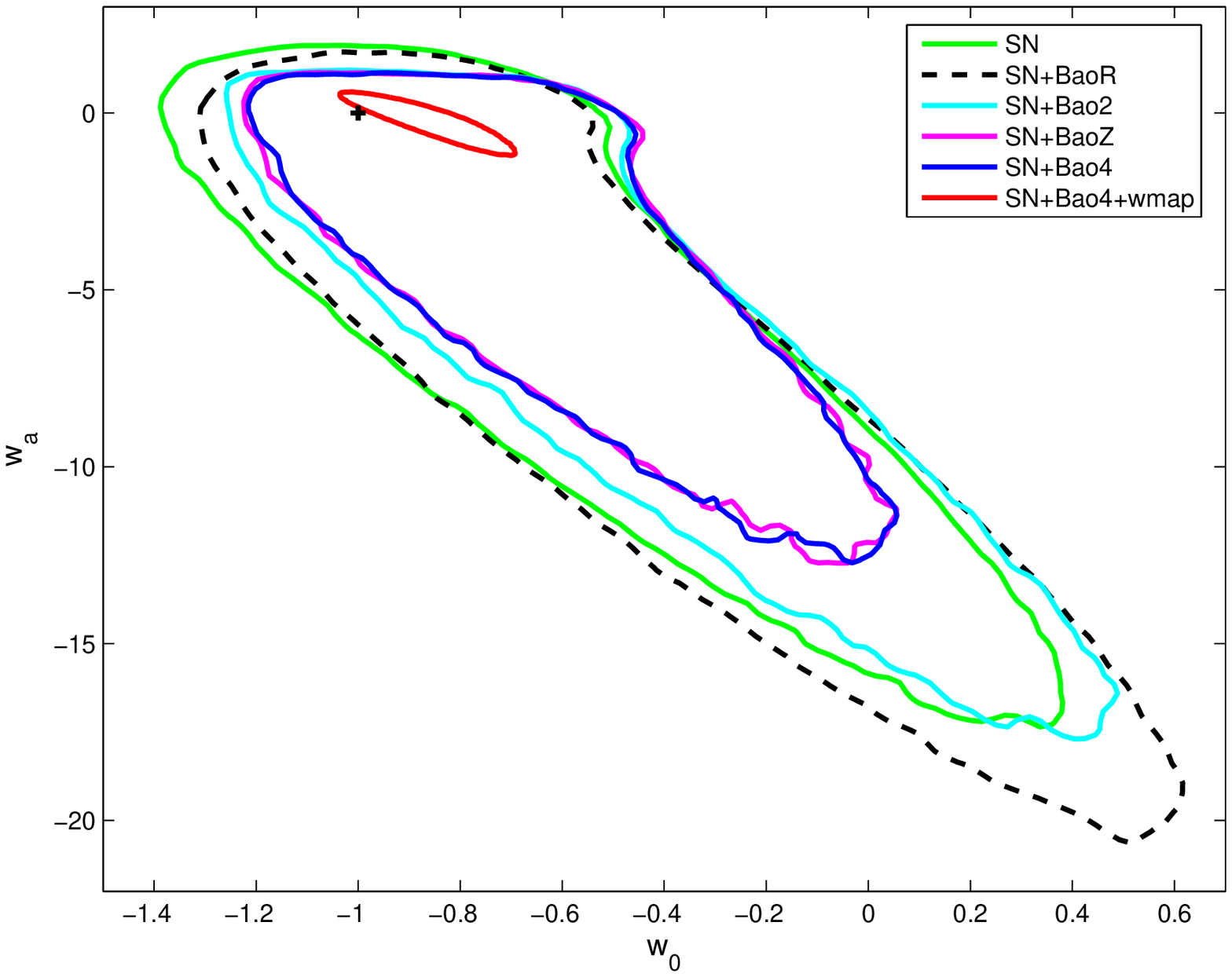}
\label{cstadcpl:c}}
\subfigure[]{ \includegraphics[width=3in]{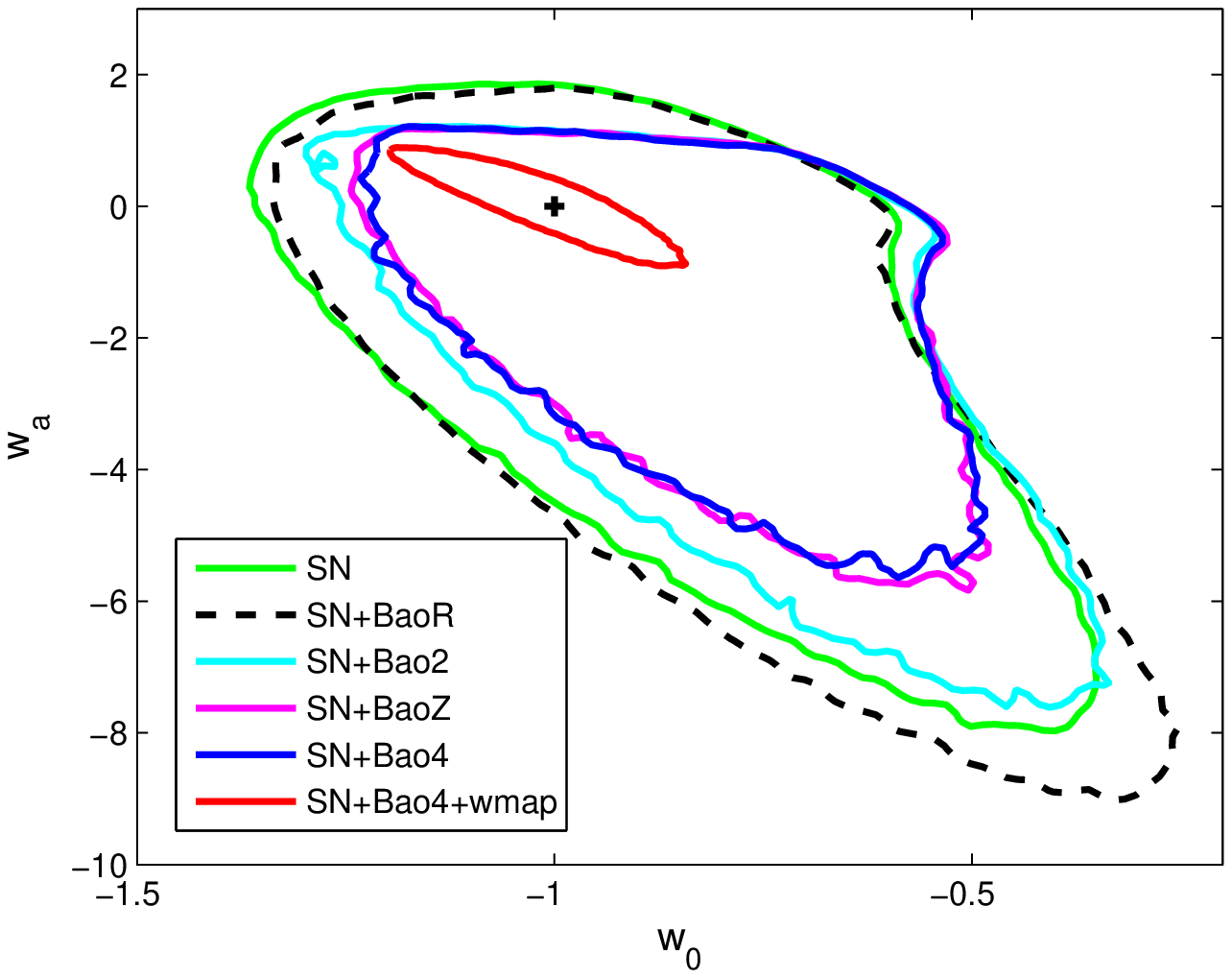}
\label{cstadcpl:d}}
\end{array}$
\caption{The marginalized $1\sigma$ contours of $w_0$ and $w_a$ in the CPL model. The green line is for SNIa,
the dashed black line is for SNIa+BaoR, the cyan line is
for SNIa+Bao2, the magenta line is for SNIa+BaoZ, the blue line is for SNIa+Bao4, and the red line is for SNIa+Bao4+WMAP5.
The SNIa used in \subref{cstadcpl:a}-\subref{cstadcpl:d} are Csta, Cstb, Cstc, and Cstd, respectively.} \label{cstadcpl}
\end{figure}

\begin{figure}[t]
$\begin{array}{c}
\subfigure[]{\includegraphics[width=3in]{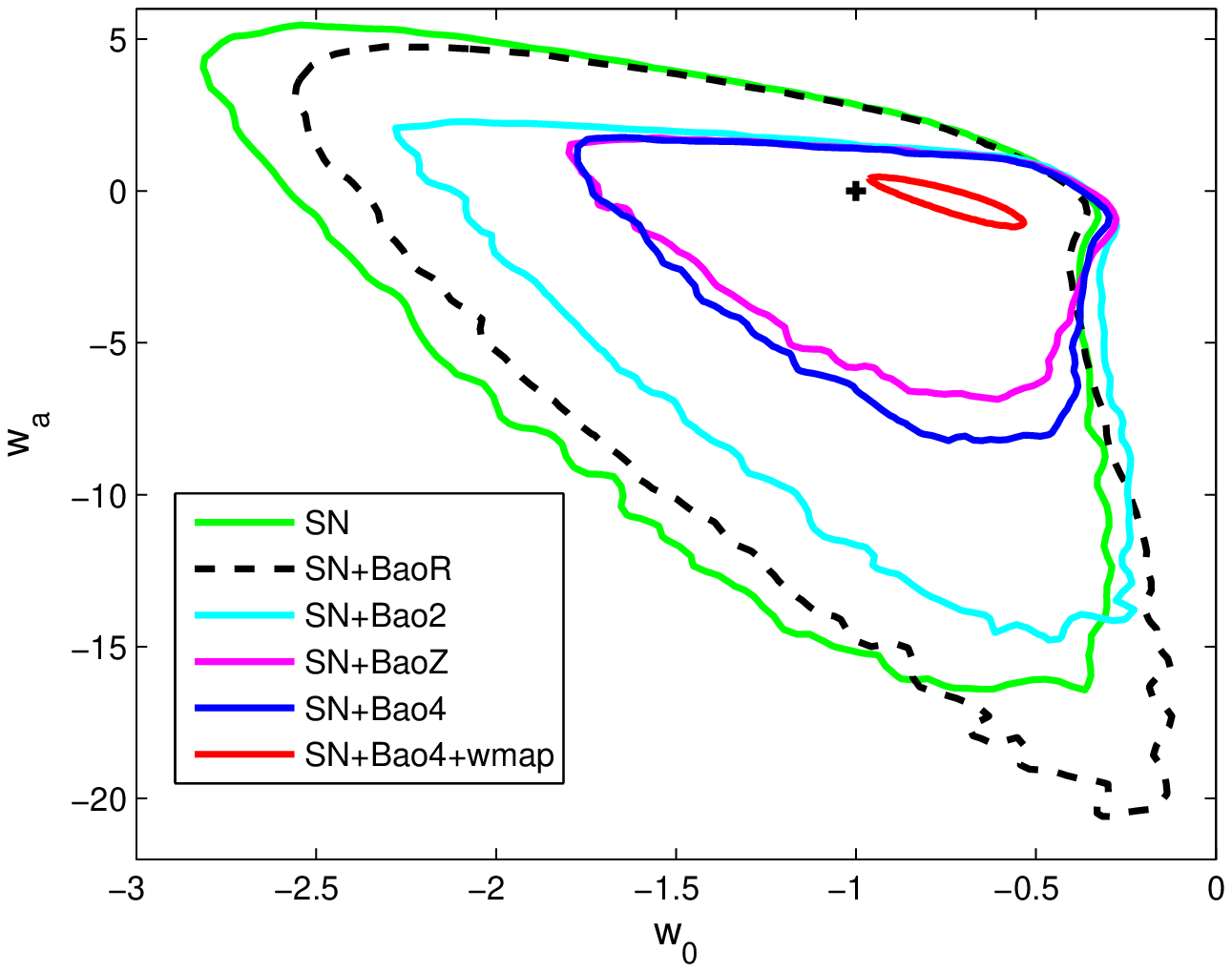}
\label{sd2fits:a}}
\subfigure[]{ \includegraphics[width=3in]{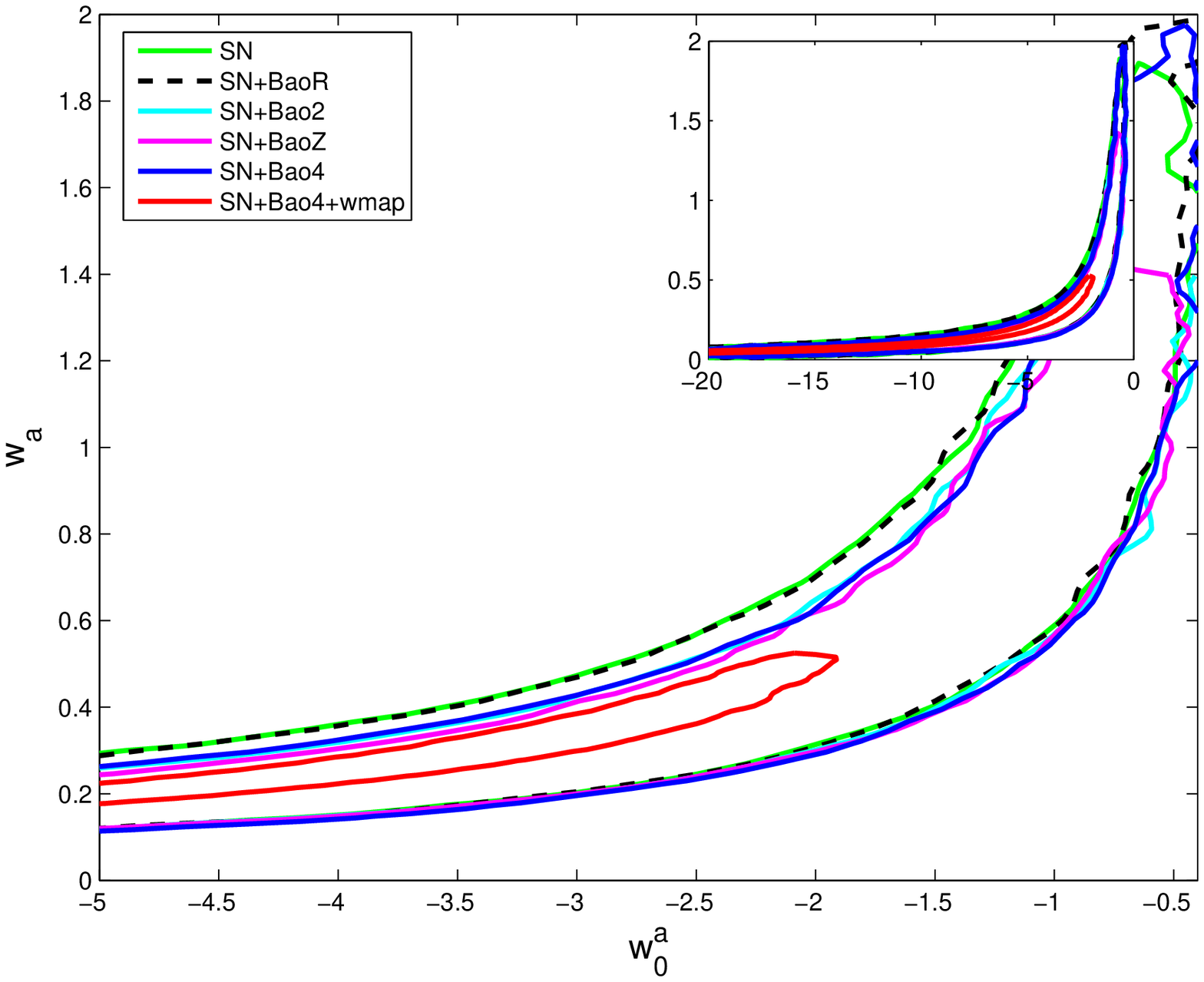}
\label{sd2fits:b}}\\
\subfigure[]{\includegraphics[width=3in]{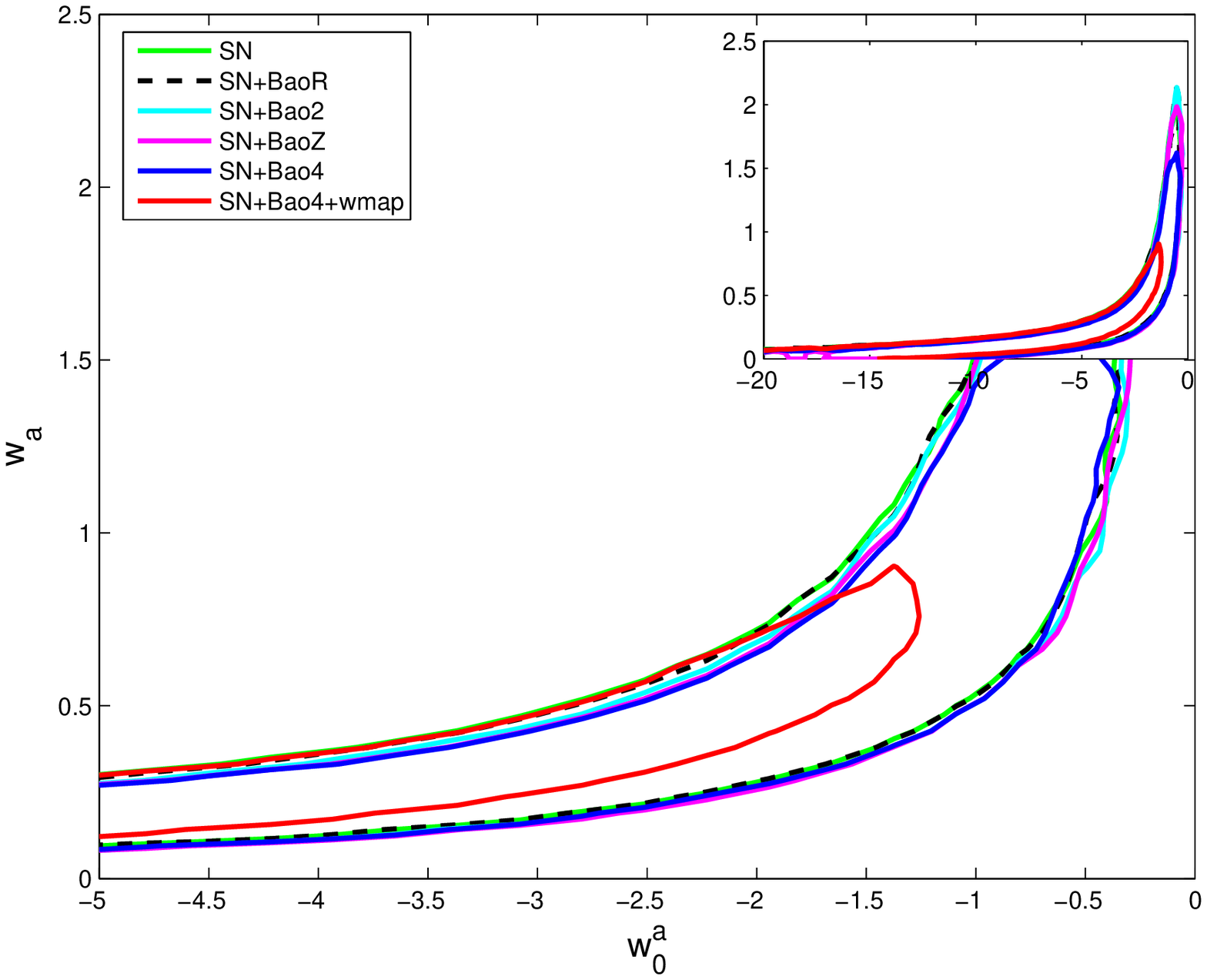}
\label{sd2fits:c}}
\subfigure[]{ \includegraphics[width=3in]{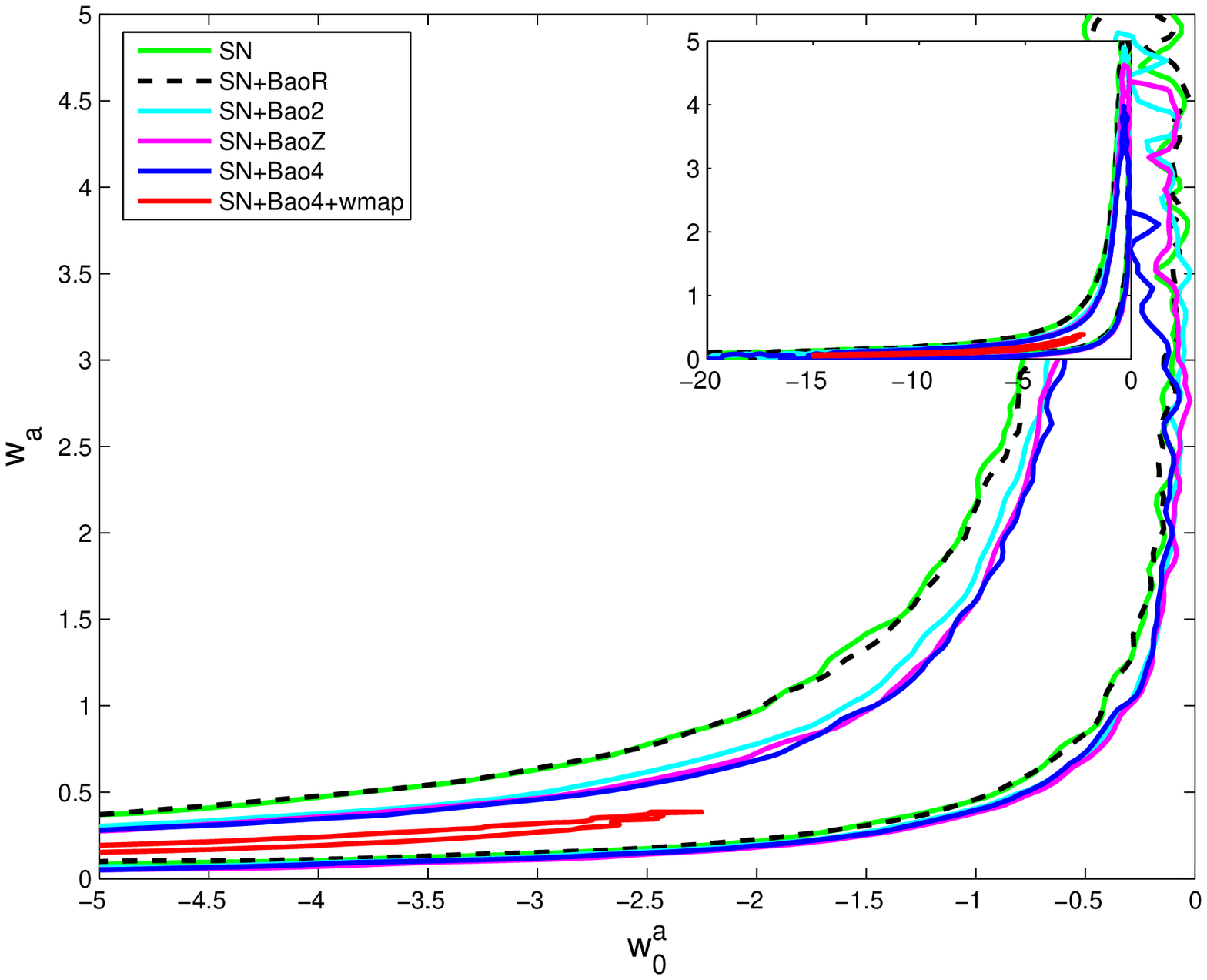}
\label{sd2fits:d}}
\end{array}$
\caption{The marginalized $1\sigma$ contours of $w_0$ ($w_0^a$) and $w_a$. The label of contours is the same as Fig. \ref{cstadcpl}.
\subref{sd2fits:a} is for the CPL model with Sdss2 SNIa data. \subref{sd2fits:b}-\subref{sd2fits:d} are for the Wetterich model.
The SNIa used in \subref{sd2fits:b}-\subref{sd2fits:d} are Csta, Cstd and Sdss2, respectively.} \label{sd2fits}
\end{figure}

Fig. 1 shows contours of the fitting results for the CPL parametrization by combining different SNIa data together
with different BAO data and
the combination of the WMAP5 data. The SNIa data used in Figs. 1a-1d and 2a are Csta, Cstb, Cstc, Cstd, and Sdss2, respectively.
The green lines are for using
just SNIa data alone with different templates. The dashed black lines are for SN+BaoR, the cyan lines are for SN+Bao2.
 Since the BaoR data is
derived from Bao2 data, so the  result using BaoR data is compatible with that of Bao2. Also we can see that the
constraint is a little better
with Bao2 than that with BaoR. The magenta lines are for SN+BaoZ, we see that the constraint from BaoZ is
in general consistent with
that from Bao2 although the former gives a little tighter constraint. These behaviors keep the same for SNIa data
 with different templates. The
blue lines are for SN+Bao4, where Bao4 is the combination of Bao2+BaoZ.
The combination of the SN+Bao4+WMAP5
is  shown in the red solid lines. The
observation that the CPL ansatz is strained to describe the DE behavior suggested by data at low and high redshifts
by comparing Csta+BaoR and
Csta+BaoR+CMB shift parameter \cite{star} has been reduced by using the same SN data (Csta) with BaoZ or Bao4,
or other SN data sets (Cstb-Cstd, Sdss2) with
arbitrary combinations of BAO data. This confirms that the systematics in some of the data sets really matters the
fitting results.

\begin{figure}[t]
\includegraphics[width=6in]{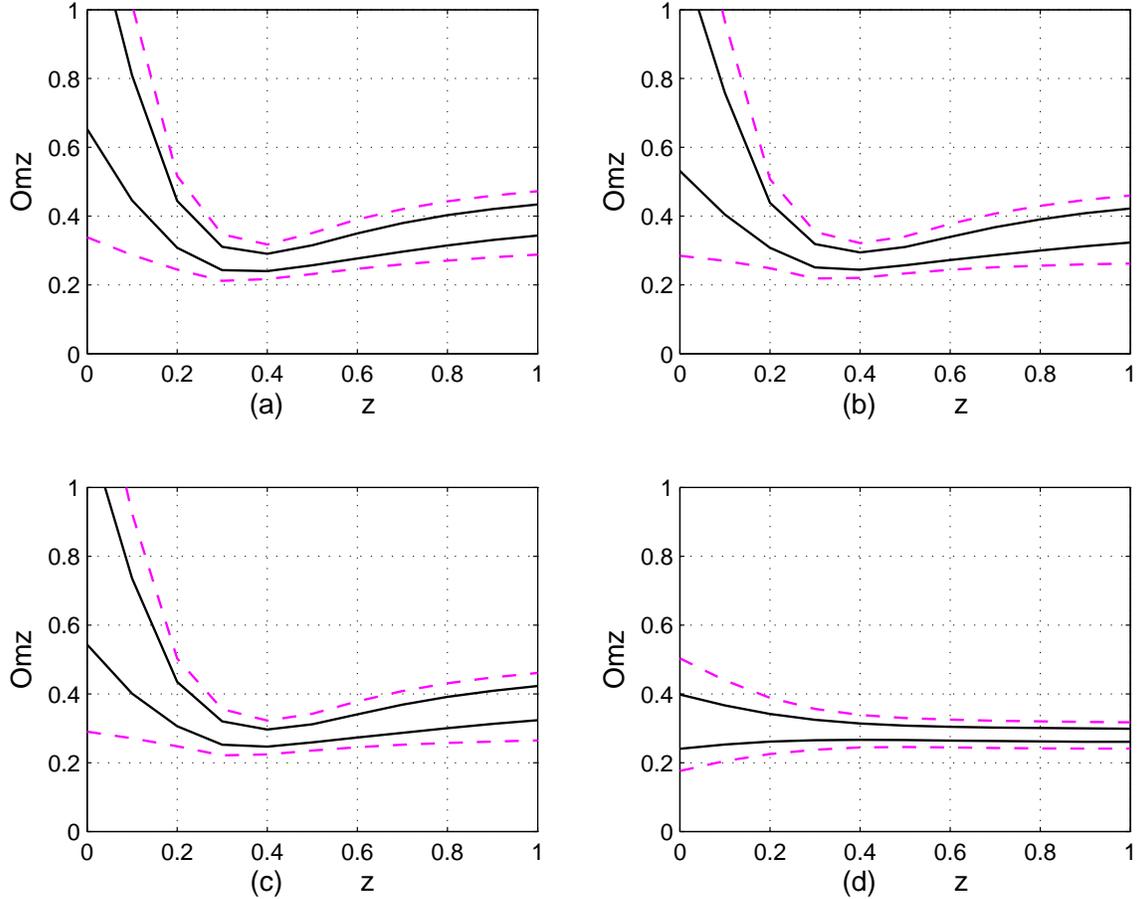}
\caption{The marginalized $1\sigma$ and $2\sigma$ errors of $Om(z)$ reconstructed in the CPL model
by using the Csta SNIa data.
(a) uses the combination of SNIa+BaoR, (b) uses the combination of SNIa+BaoZ,
(c) uses the combination of SNIa+Bao4
and (d) uses the combination of SNIa+Bao4+WMAP5.} \label{cstaomzcpl}
\end{figure}

\begin{figure}[t]
\includegraphics[width=6in]{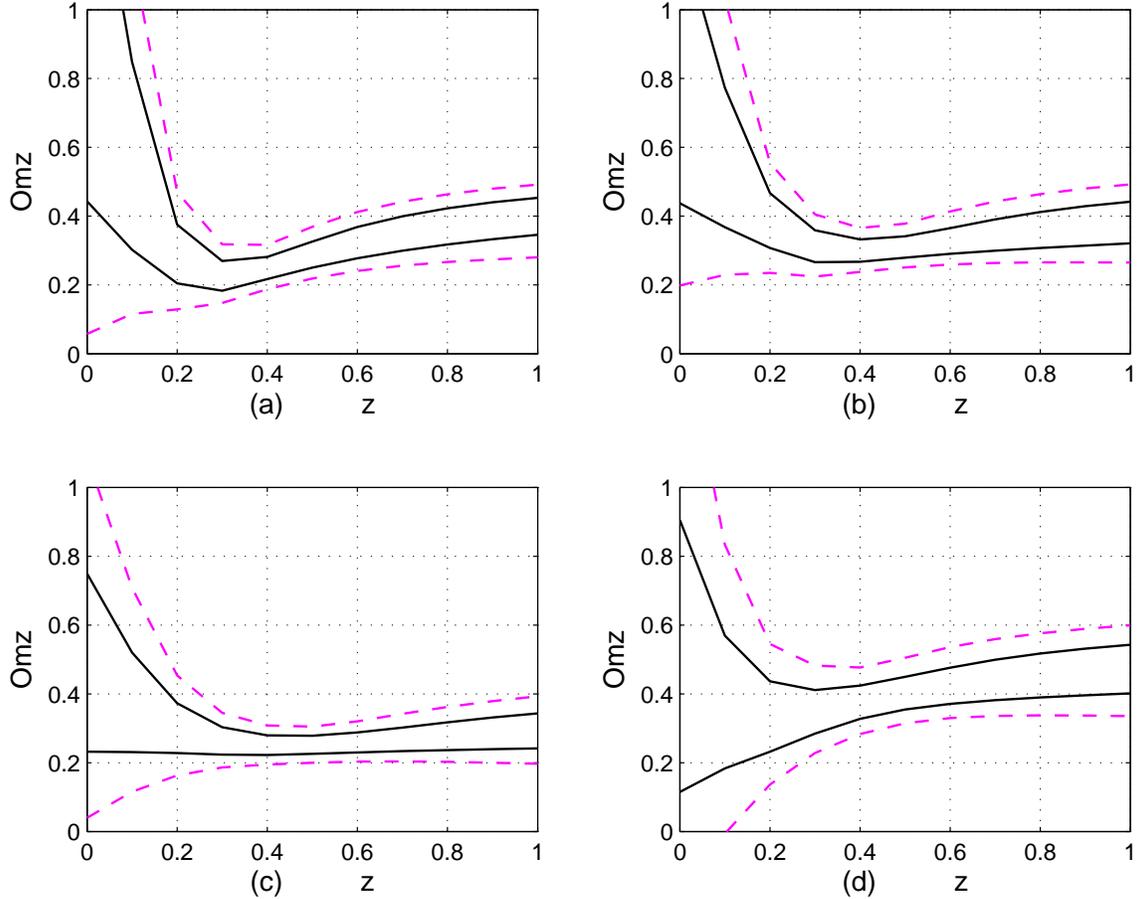}
\caption{The marginalized $1\sigma$ and $2\sigma$ errors of $Om(z)$ reconstructed in the CPL model with SNIa+BaoR.
The SNIa data used in (a)-(d) are Cstb, Cstc, Cstd, and Sdss2, respectively.} \label{snbrcpl}
\end{figure}

For the Csta data, the reconstructed $Om(z)$ for SNIa data in combination with BAO and CMB data is shown in Fig. \ref{cstaomzcpl}.
Since $Om(z)$ is not a constant at $1\sigma$ level if we combine Cata SNIa with BaoR, BaoZ or Bao4,
so the $\Lambda$CDM model is excluded at $1\sigma$ level if use the combination of Csta SNIa and BAO data.
However, when the WMAP5 data is added, the $\Lambda$CDM model is consistent at $1\sigma$ level.
The growth in the value of $Om(z)$ at low redshifts becomes smaller when we change the BAO data
from BaoR to BaoZ or Bao4.
In Fig. \ref{snbrcpl}, we show the marginalized $1\sigma$ and $2\sigma$ errors of $Om(z)$ reconstructed using the BaoR and different
SNIa data sets. The growth
in the value of $Om(z)$ at low redshifts in Fig. 3a by using
Csta+BaoR as observed in \cite{star} gets flattened if we change
SNIa data sets from Csta to Cstb, Cstc, Cstd and Sdss2. This shows
that the finding in \cite{star} is not a general behavior even at
low $z$. The evolutional behavior of $Om(z)$ at low redshifts is
changed for different selections of the SNIa data.
Thus the striking observation that our universe
is slowing down \cite{star} based on the fitting at low redshifts
for CPL ansatz is caused by the systematics of the specially chosen
SNIa and BAO data sets. By choosing some other data sets, the CPL
parametrization is in compatible with combinations of data sets at
low and high redshifts. The un-evolving $Om(z)$ is also allowed even
at low redshifts. From Figs. 3a and 4, we see that the $\Lambda$CDM model is
not allowed at $1\sigma$ level for the combination of BaoR and Csta, Cstb or Cstc,
although the cosmological constant is inside the $1\sigma$ contours of $w_0$
and $w_a$ in Figs. 1b and 1c. This shows the advantage of $Om$ diagnostic because the
uncertainty of $\Omega_{m0}$ is taken accounted for.

In the above discussions we have focused on the commonly used
functional form  for DE, the CPL parametrization, and argued that
the systematics in SNIa and BAO data sets really affects the
evolution of the DE and the acceleration of the universe. Besides
the systematics of data sets, whether the influences on the
evolution of the DE and the acceleration are caused by the specific
choice of the parametrization of the DE is another interesting
question to ask. In \cite{star}, the incompatibility of the CPL
ansatz fitting to the data simultaneously at low and high redshifts was
attributed to the versatility of the CPL parametrization. Choosing
another ansatz of DE parametrization, it was found significantly
better than that of CPL, that ansatz  can provide a good fit to the
combination of Csta+BaoR and the CMB shift data. In the rest of this
work we are going to further examine the versatility of the CPL
parametrization by considering a different ansatz of DE
parametrization and confronting it with different combinations of
various data sets as used above.

What we are going to study is the Wetterich parametrization of the
form \cite{par1}
\begin{equation}
\label{wetwz}
w(z)=\frac{w_0}{1+w_a \ln(1+z)}.
\end{equation}
When $z=0$, $w(z)=w_0$, and at large redshift $z\gg 1$, $w(z)\approx
0$. The Wetterich parametrization becomes the constant
parametrization $w(z)=w_0$ when $w_a=0$.

In this model, the dimensionless Hubble parameter for a flat
universe is
\begin{equation}
\label{wetez}
E^2(z)=\Omega_{m0}(1+z)^3+(1-\Omega_{m0})(1+z)^3[1+w_a\ln(1+z)]^{3w_0/w_a}.
\end{equation}
To ensure the positivity of DE, we require that $w_a\ge 0$. For the convenience of
numerical calculation, we take the independent parameters as $w_0^a=w_0/w_a$ and $w_a$. The
parameters in this model are $\bm{p}=(\Omega_{m0},\ w_0^a,\ w_a)$, the number is the
same as those in the case of CPL ansatz.

\begin{figure}[t]
\includegraphics[width=6in]{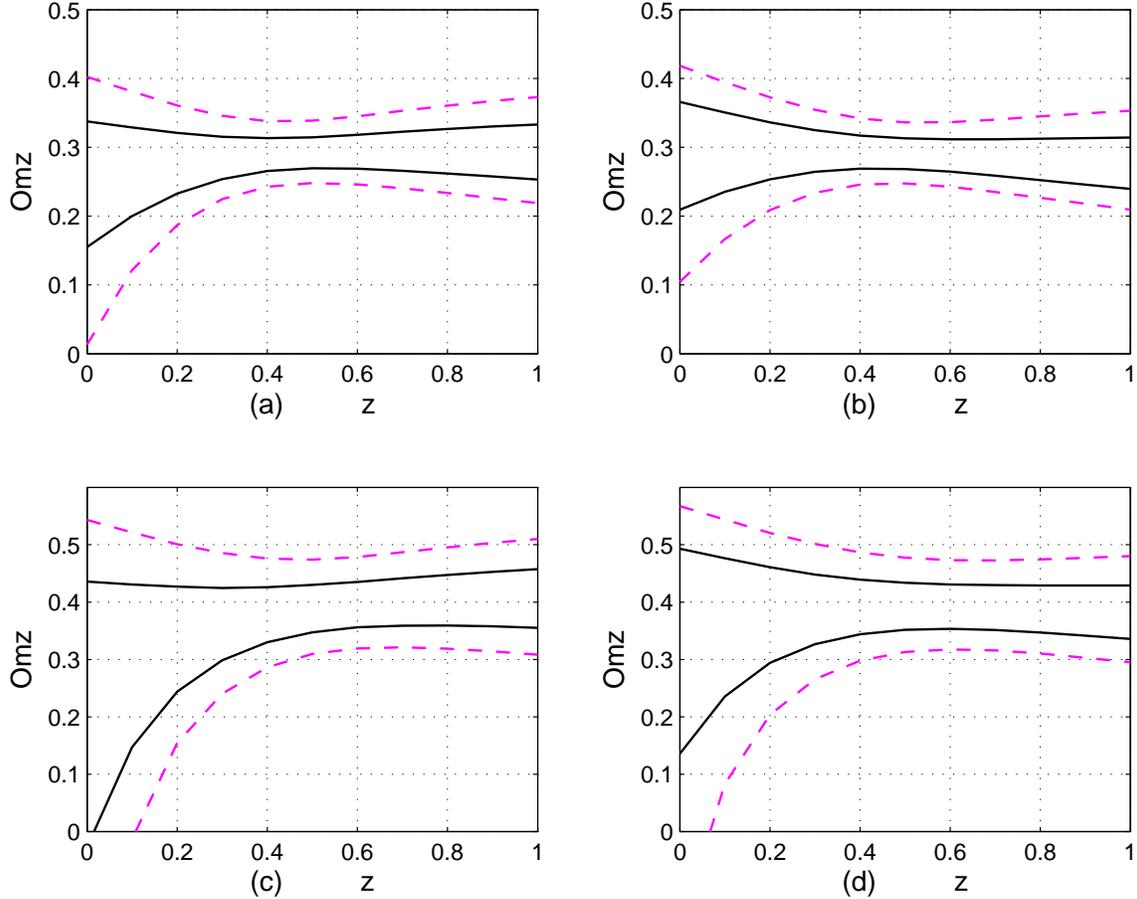}
\caption{The marginalized $1\sigma$ and $2\sigma$ errors of $Om(z)$ reconstructed in the Wetterich model.
(a) uses the combination of Csta SNIa+BaoR, (b) uses the combination of Csta SNIa+BaoZ, (c) uses the combination of Sdss2 SNIa+BaoR,
and (d) uses the combination of Sdss2 SNIa+BaoZ.} \label{wetomz}
\end{figure}

The compatibility checks in confronting with different SNIa, BAO
data sets together with CMB data are shown in  Figs. 2b-2d. In Figs.
2b-2d, we have used the Csta, Cstd and Sdss2 SNIa data sets,
respectively. As indicated in Fig. 1 and Fig. 2, the green lines are
just for SNIa data alone, the dashed black lines are for SN+BaoR, the
cyan lines are results for SNIa+Bao2, the magenta lines indicate
SNIa+BaoZ, the blue lines are for SNIa+Bao4, and the solid red lines are for the combination of
SNIa+Bao4+WMAP5. It is interesting to see that different from the CPL
ansatz, this parametrization allows the compatibility of different
SNIa, Bao and CMB data sets, even for comparing the combinations of
Csta+BaoR, Csta+BaoZ and Csta+Bao4+WMAP5. The Wetterich
ansatz is found able to fit different data sets both at low and high
redshifts. In Fig. \ref{wetomz}, we  present the behaviors of the $Om(z)$
reconstructed by comparing the influence of different SNIa and BAO
data sets. Interestingly, the evolution of $Om(z)$ is not affected
much by different SNIa and BAO data sets. The $Om(z)$ is perfectly
consistent with $\Lambda$CDM for different data sets.

For the comparison of different data sets and different parametrizations, we
summarize the minimum value of $\chi^2$
in Table \ref{table1}. From Table \ref{table1}, we observe the
reliability of parametrizations when using
different data sets. Comparing two different parametrizations, we see that
both of them can fit well of the
data, while the CPL model is a little better than the Wetterich
parametrization for the Csta SNIa and the combination
of Csta SNIa with BAO data.

\begin{table}[htp]
\begin{center}
\caption{The minimum value of $\chi^2$ for different combinations of data sets and models.} \label{table1}
\begin{tabular}{|l|c|c|}
\hline Data  & CPL ($\chi^2$/DOF) & Wetterich \\ \hline
Csta & 462.06/394 & 466.33/394\\
Csta+BaoR & 462.43/395 & 467.62/395\\
Csta+Bao2 & 462.45/394 & 467.73/394\\
Csta+BaoZ & 462.10/394 & 466.60/394\\
Csta+Bao4 & 464.11/396 & 468.57/396\\
Csta+Bao4+WMAP5 & 468.73/399 & 468.69/399\\ \hline
Sdss2 & 227.55/285 & 227.09/285\\
Sdss2+BaoR & 229.77/286 & 229.45/286\\
Sdss2+Bao2 & 229.93/285 & 229.65/285\\
Sdss2+BaoZ & 228.77/285 & 228.57/285\\
Sdss2+Bao4 & 231.15/287 & 231.02/287\\
Sdss2+Bao4+WMAP5 & 231.71/290 & 231.94/290
 \\ \hline
\end{tabular}
\end{center}
\end{table}

\section{Conclusion}
To summarize, we have examined the influence of the systematics in different data sets in SNIa and BAO on the
 fitting results of the CPL
parametrization. We found that the tension observed in \cite{star} between low $z$ (Csta+BaoR) and the high
 $z$ (CMB) data is not a general
behavior. By using SNIa with other templates and other BAO data
sets, the incompatibility of the CPL parametrization will disappear.
This result supports the speculation that the systematics in the
data sets can affect the fitting results and leads to different
evolution of the DE model \cite{star}. However this answer is still
not definite. The different evolutions in $Om(z)$ as we observed by
using different combinations of various SNIa and BAO data sets
disappear when we use the Wetterich parametrization to replace the
CPL parametrization. This again brings the problem as raised in
\cite{star} that the CPL parametrization might be blamed to be not
so versatile. In order to disclose
 the exact answer, we have to
examine more models of DE. This work offers the attempt of using different combinations of various data sets at
low and high redshifts to examine
the  effects of data systematics on the evolution of the DE and the acceleration of the universe. Further investigation
along this line by
including more data sets and examining more DE models are called for and we will report our progress in this direction
in the future work.

\begin{acknowledgments}
RGC and BW thank Chongqing University of Posts and
Telecommunications for the warm hospitality during their visits.
This work was partially supported by NNSF of China (Nos. 10821504, 10878001, 10975168 and 10935013)
and the National Basic Research Program of China under
grant No. 2010CB833004.
YG was partially supported by the Natural Science Foundation Project of CQ CSTC under
grant No. 2009BA4050.

\end{acknowledgments}

\end{document}